\documentclass[aip,preprint]{revtex4-1}
\pdfoutput=1
\draft

\usepackage{graphicx}
\usepackage{bm}

\begin{document}

\title{Quantum oscillations in diamond field effect transistors with a $h$-BN gate dielectric} 
\author{Yosuke Sasama}
\affiliation{International Center for Materials Nanoarchitectonics, National Institute for Materials Science, Tsukuba 305-0044, Japan}
\affiliation{University of Tsukuba, Tsukuba, 305-8571, Japan}

\author{Katsuyoshi Komatsu}
\affiliation{International Center for Materials Nanoarchitectonics, National Institute for Materials Science, Tsukuba 305-0044, Japan}

\author{Satoshi Moriyama}
\affiliation{International Center for Materials Nanoarchitectonics, National Institute for Materials Science, Tsukuba 305-0044, Japan}

\author{Masataka Imura}
\affiliation{Research Center for Functional Materials, National Institute for Materials Science, Namiki, Tsukuba 305-0044, Japan}

\author{\\Shiori Sugiura}
\affiliation{Research Center for Functional Materials, National Institute for Materials Science, Sakura, Tsukuba 305-0003, Japan}

\author{Taichi Terashima}
\affiliation{Research Center for Functional Materials, National Institute for Materials Science, Sakura, Tsukuba 305-0003, Japan}

\author{Shinya Uji}
\affiliation{Research Center for Functional Materials, National Institute for Materials Science, Sakura, Tsukuba 305-0003, Japan}
\affiliation{University of Tsukuba, Tsukuba, 305-8571, Japan}

\author{Kenji Watanabe}
\affiliation{Research Center for Functional Materials, National Institute for Materials Science, Namiki, Tsukuba 305-0044, Japan}

\author{Takashi Taniguchi}
\affiliation{Research Center for Functional Materials, National Institute for Materials Science, Namiki, Tsukuba 305-0044, Japan}

\author{Takashi Uchihashi}
\affiliation{International Center for Materials Nanoarchitectonics, National Institute for Materials Science, Tsukuba 305-0044, Japan}

\author{Yamaguchi Takahide}
\email[]{YAMAGUCHI.Takahide@nims.go.jp}
\affiliation{International Center for Materials Nanoarchitectonics, National Institute for Materials Science, Tsukuba 305-0044, Japan}
\affiliation{University of Tsukuba, Tsukuba, 305-8571, Japan}

\date{\today}

\begin{abstract}
Diamond has attracted attention as a next-generation semiconductor because of its various exceptional properties such as a wide bandgap and high breakdown electric field. Diamond field effect transistors, for example, have been extensively investigated for high-power and high-frequency electronic applications. The quality of their charge transport (i.e., mobility), however, has been limited due to charged impurities near the diamond surface. Here, we fabricate diamond field effect transistors by using a monocrystalline hexagonal boron nitride as a gate dielectric. The resulting high mobility of charge carriers allows us to observe quantum oscillations in both the longitudinal and Hall resistivities. The oscillations provide important information on the fundamental properties of the charge carriers, such as effective mass, lifetime, and dimensionality. Our results indicate the presence of a high-quality two-dimensional hole gas at the diamond surface and thus pave the way for studies of quantum transport in diamond and the development of low-loss and high-speed devices.
\end{abstract}

\maketitle

Shubnikov-de Haas (SdH) oscillations are a representative quantum transport phenomenon that is caused by Landau quantization of electronic states under an applied magnetic field.\cite{Sho84} The frequency of the oscillation corresponds to the extremal cross-sectional area of the Fermi surface perpendicular to the magnetic field, whereas the temperature and magnetic-field dependences of the oscillation allow one to estimate the effective mass and lifetime of the charge carriers. The Berry phase associated with the cyclotron orbit and Land\'e $g$ factor can also be estimated. Thus, SdH oscillations are a powerful probe for investigating the fundamental properties of electronic states. One of the systems in which SdH oscillations have been intensively studied is the two-dimensional electron and hole gas in semiconductor field-effect transistors (FETs); the semiconductors include emerging two-dimensional materials like graphene\cite{Nov04}, transition-metal dichalcogenides\cite{Cui15} and black phosphorus\cite{Li15}, as well as conventional Si\cite{Fow66,Kli74}, Ge\cite{Bin79}, GaAs\cite{Sto80,Col89} and wide-band-gap GaN\cite{Kha92}, ZnO\cite{Tsu07}, Ga$_2$O$_3$\cite{Zha18}.

Diamond, with the same crystal structure as Si and Ge, is a promising future semiconductor material due to its excellent properties, such as a wide bandgap, high breakdown electric field, high thermal conductivity, and high intrinsic mobilities\cite{Wor08}. These properties are all superior to those of Si and Ge\cite{Sze02}. There have been extensive studies on diamond FETs\cite{Kaw14,Gul14,Var14,Rus15,Mat16,Kas17,Ren17,Liu17,Pha17,Kar17,Wan19} aimed at developing applications for high-power and high-frequency electronics. SdH oscillations in diamond, however, have been difficult to observe because of the low mobilities of the charge carriers in FET structures and have been reported only in our previous paper\cite{Tak14}. In that study, the SdH oscillations were observed in ionic-liquid-gated diamond FETs. Although the SdH oscillations provided evidence of the two-dimensional character of the hole gas that accumulates at the diamond surface, no clear dependence of the oscillations on gate voltage was observed, and the indicated spatial inhomogeneity of the hole gas remained an issue to be resolved.

In this study, we create high-mobility two-dimensional hole gas in diamond using monocrystalline hexagonal boron nitride ($h$-BN) as a gate dielectric of diamond FETs [Figs. 1(a) and 1(b)]. SdH oscillations are detected in both longitudinal ($\rho_{xx}$) and Hall ($\rho_{yx}$) resistivities with their frequencies clearly dependent on the gate voltage. The oscillations as well as the background magnetic-field dependence of $\rho_{xx}$ and $\rho_{yx}$ are quantitatively explained by the standard theory of SdH oscillations. This result, combined with the suggested homogeneity of the electronic system, indicates the high quality of the hole gas. The estimated quantum lifetime also provides useful insights into the carrier scattering, which limits the mobility of the hole gas.

\begin{figure}
\includegraphics[width=15truecm]{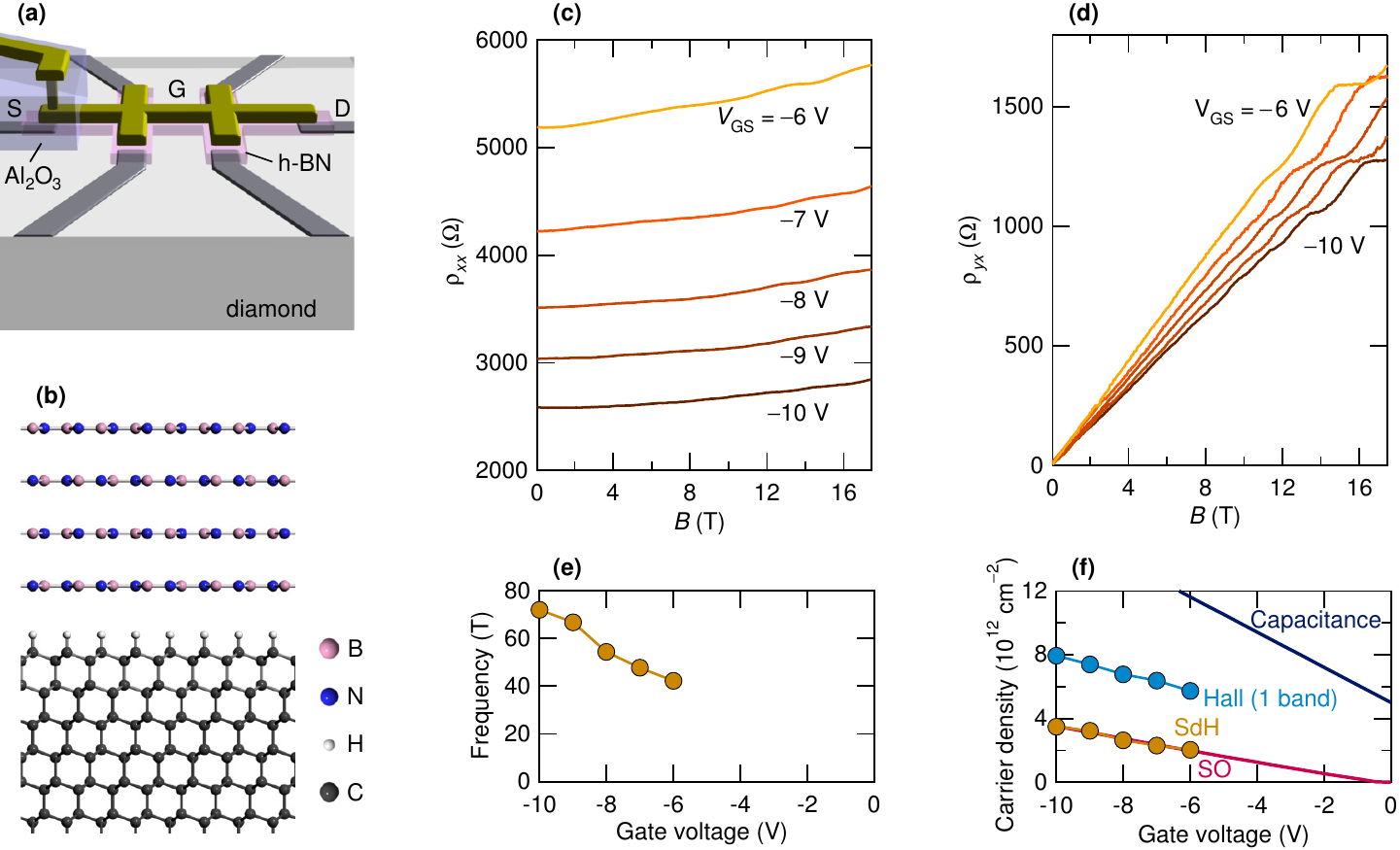}
\caption{\textbf{Shubnikov-de Haas oscillations in diamond FET with a monocrystalline $h$-BN gate dielectric.} (a) Schematic diagram of a diamond FET with a $h$-BN gate dielectric. The diamond surface in the region covered with the $h$-BN is hydrogen terminated for $p$-type conductivity, whereas the surface in the other region is oxygen terminated for electrical isolation. Al$_2$O$_3$ is shown in a limited region in this figure for clarity, but it covers the entire diamond surface in the actual devices. (b) Schematic diagram of $h$-BN/hydrogen-terminated diamond heterostructure. (c) and (d) Longitudinal (c; $\rho_{xx}$) and Hall (d; $\rho_{yx}$) resistivities of device S1 at different gate voltages at 30 mK plotted as a function of magnetic field. Shubnikov-de Haas oscillations are seen in both $\rho_{xx}$ and $\rho_{yx}$. (e) Frequency of the SdH oscillation in $\sigma_{xy}(=\rho_{yx}/(\rho_{xx}^2+\rho_{yx}^2)$), which depends monotonically on gate voltage. (f) Gate voltage dependence of the sheet carrier densities calculated from the SdH frequency and low-field Hall resistivity. The figure also shows the carrier density estimated from the gate capacitance and the simulated density of the holes occupying the split-off (SO) hole subband.}
\end{figure}

The fabrication process and the electrical properties of the devices at room temperature have been reported in detail elsewhere\cite{Sas18}. Briefly, a monocrystalline $h$-BN was cleaved by using Scotch tape and laminated\cite{Dea10} on a hydrogen-terminated diamond surface, which acted as a channel of FETs. The hydrogen-terminated diamond surface is stable in air, has a low density of dangling bonds, and favors the accumulation of holes due to the energetically high valence band\cite{Mai00,Squ06}. Unprecedentedly high mobilities $\textgreater$300 cm$^2$V$^{-1}$s$^{-1}$ were obtained for the holes at room temperature, thanks to the excellent properties of $h$-BN as a dielectric. For the carrier densities exceeding $n_c=(4-6){\times}10^{12}$ cm$^{-2}$, the channel remains conductive at low temperatures.\cite{Sas18} (See also Fig. S1 of the Supplemental Material.) At lower carrier densities, the resistance increases with decreasing temperature, which is attributed to localization of the hole carriers. The carrier density $n_c$ required to make the holes conductive at low temperature in $h$-BN-gated diamond FETs is an order of magnitude smaller than that in ionic-liquid-gated diamond FETs in our previous studies\cite{Yam13,Tak18}. This suggests that the hole gas in the $h$-BN-gated FETs was of higher quality. Experimental results for two devices are described in this paper. The Hall mobility at low temperature reached $304$ cm$^2$V$^{-1}$s$^{-1}$ at a Hall carrier density $7.9\times10^{12}$ cm$^{-2}$ in device S1 and $410$ cm$^2$V$^{-1}$s$^{-1}$ at $5.9\times10^{12}$ cm$^{-2}$ in device S2. These values are higher than those (52-91 cm$^2$V$^{-1}$s$^{-1}$) in ionic-liquid-gated diamond FETs.

The magnetic-field dependences of $\rho_{xx}$ and $\rho_{yx}$ measured at low temperature for $n_\mathrm{Hall}{\textgreater}n_c$ shows SdH oscillations [Fig. 1(c) and 1(d)]. The gate voltage dependence of the oscillation is reproducible. The magnetic-field dependences of $\rho_{xx}$ and $\rho_{yx}$ were nearly unchanged after the gate voltage was swept back to zero and swept again to the same voltage. Furthermore, the $\rho_{yx}$ curves obtained using two different sets of Hall probes were in good agreement (Fig. S2 of the Supplemental Material), and the oscillation in $\rho_{yx}$ was consistent with the oscillation in $\rho_{xx}$ (see below). The frequency $B_F$ of the oscillation monotonically increased with increasing negative gate voltage. [Fig. 1(e)] These results indicate stable and uniform controllability of the hole gas with the gate voltage in the present FETs.

A notable feature in Fig. 1(d) is that $\rho_{yx}(B)$ has a steplike structure. This is reminiscent of quantum Hall plateaus, but the present system is unlikely in the quantum Hall regime. $\rho_{xx}(B)$ does not have any corresponding dips at magnetic field values where the plateau-like structures appear in $\rho_{yx}(B)$. The oscillations in $\rho_{xx}$ and $\rho_{yx}$ would be in quadrature if the system was in the quantum Hall regime, but the oscillations in $\rho_{xx}$ and $\rho_{yx}$ are nearly in antiphase. In fact, the oscillations in $\rho_{yx}$ can be understood as normal SdH oscillations as shown below. Observation of the quantum Hall effect may be achievable upon further improvement of the quality of the diamond/$h$-BN heterostructure.

\begin{figure}
\includegraphics[width=6.5truecm]{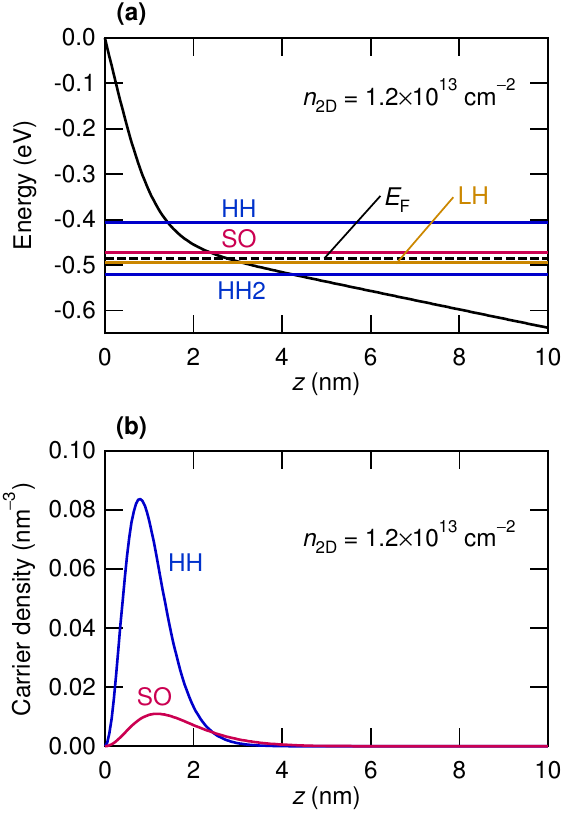}
\caption{\textbf{Simulated subband structure at the diamond (111) surface.} (a) Energies at the top of subbands obtained by solving the Schr\"odinger and Poisson equations in a self-consistent manner. The figure also shows the Fermi level ($E_\mathrm{F}$)) and the spatial variation of potential energy.
$z$ is the depth from the diamond surface. The total hole density is $1.2\times10^{13}$ cm$^{-2}$. The band mixing of the heavy-, light-, and split-off holes is neglected in the calculation. HH, LH, and SO represent the lowest subbands of heavy-, light- and split-off holes, and HH2 represents the second subband of heavy holes. (b) Hole density profiles of the lowest heavy- and lowest split-off-hole subbands for total carrier density of $1.2\times10^{13}$ cm$^{-2}$.}
\end{figure}

The frequency of the SdH oscillations provides an estimate of the carrier density through the relation $n_\mathrm{SdH}=(2e/h)B_F$. The carrier density $n_\mathrm{SdH}$ thus obtained is plotted as a function of gate voltage in Fig. 1f. Here, $n_\mathrm{SdH}$ differs substantially from $n_\mathrm{Hall}$, which is estimated from the low-field Hall coefficient by assuming a single-carrier model. A similar discrepancy between $n_\mathrm{SdH}$ and $n_\mathrm{Hall}$ was observed in the other devices (Fig. S3d of the Supplemental Material). Such discrepancies or missing carriers have been observed in other heterointerface systems\cite{Bin79,Sto80,Cav10} and in ionic-liquid-gated diamond FETs\cite{Tak14}. They are generally attributed to the existence of parallel conduction channels with low carrier mobilities and/or spatial inhomogeneity in the electronic system. 

\begin{figure}
\includegraphics[width=7.5truecm]{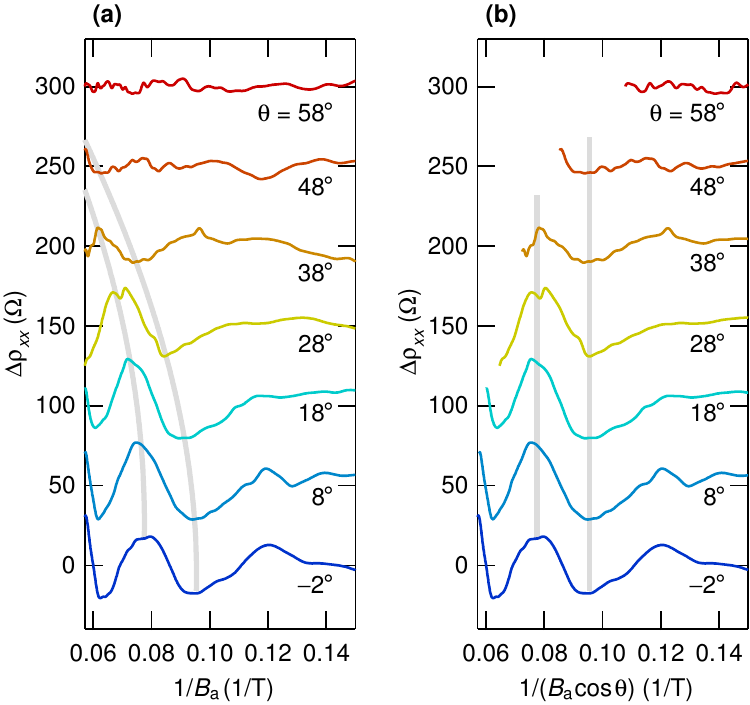}
\caption{\textbf{Shubnikov-de Haas oscillations for different magnetic field orientations.} (a) Shubnikov-de Haas oscillations observed for device S2 at $V_\mathrm{GS}=-3.5$ V for different magnetic field orientations at 30 mK. The oscillating component after subtracting the background is plotted as a function of the applied magnetic field $B_{a}$. $\theta$ is the angle between the magnetic field and the direction normal to the diamond surface. (b) The data in (a) plotted as a function of $B_{a}\cos(\theta)$, the field component perpendicular to the diamond surface. The curves in a and b are vertically offset for clarity.}
\end{figure}

The origin of the discrepancy between $n_\mathrm{SdH}$ and $n_\mathrm{Hall}$ in the present system can be specifically inferred by inspecting the valence-band electronic structure at the diamond surface. The valence bands in diamond consist of heavy-, light-, and split-off-hole bands. These bands are split into subbands at the diamond surface because the motion of the carriers along the direction normal to the surface is quantized due to the gate-induced confining potential. Our calculation based on Schr\"odinger-Poisson equations indicates that holes occupy the lowest heavy and lowest split-off subbands for the relevant range of the total carrier density expected from the gate capacitance. (Figure 2; see the Supplemental Material for details of the calculation.) The calculated density of the holes that occupy the split-off subband is close to $n_\mathrm{SdH}(V_g)$. [See Fig. 1(f).] We infer that only the holes in the split-off subband, which have a sufficiently high mobility, lead to the SdH oscillations. The holes in the heavy-hole subband, with relatively lower energies and positions closer to the surface, should have lower mobilities that are not high enough to lead to SdH oscillations. A further improvement in device quality and/or application of a higher magnetic field may allow us to observe the SdH oscillations of the holes in the heavy-hole subband as well.

The two-dimensional nature of the hole gas expected from the quantum confinement was evidenced by the oscillations in different magnetic field orientations. Figure 3(a) shows the SdH oscillations observed for different angles between the magnetic field and the direction normal to the diamond surface. The maxima and minima of the oscillations shift towards higher magnetic fields (lower $1/B$) as $\theta$ increases. When the data are plotted as a function of $B\cos(\theta)$, however, the positions of the maxima (minima) are approximately the same. [Fig. 3(b)] This indicates that the field component normal to the diamond surface leads to the oscillation and thus provides evidence that the hole gas is two dimensional. The good agreement of the SdH oscillations with the standard theory for two-dimensional systems (see below) also indicates the two-dimensional nature of the hole gas. 

\begin{figure}
\includegraphics[width=7.5truecm]{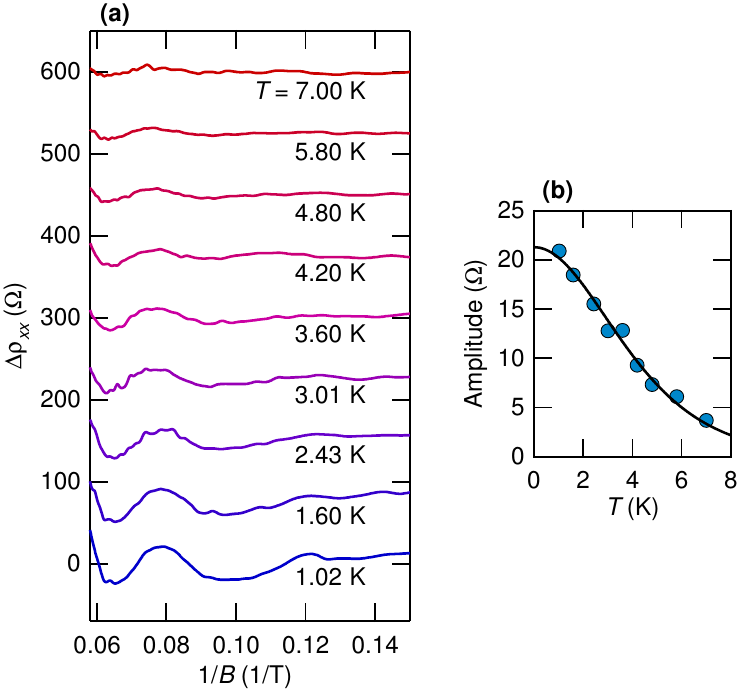}
\caption{\textbf{Shubnikov-de Haas oscillations at different temperatures.} (a) Shubnikov-de Haas oscillations observed for device S2 at $V_\mathrm{GS}=-3.5$ V at different temperatures. The curves are offset for clarity. (b) Temperature dependence of the oscillation amplitude obtained from the peak values at $1/B=0.64$ and $0.077$ (1/T). The line is a fit to the temperature reduction factor $R_T$ [Eq. (5)], providing an estimate of the effective mass $m^*/m_0=0.54\pm0.02$.}
\end{figure}

The effective mass of the carriers that show the SdH oscillations can be estimated from the temperature dependence of the oscillations. Figure 4(a) shows the oscillations at different temperatures for device S2 at a gate voltage $V_\mathrm{GS}$ of $-3.5$ V. The oscillation amplitude obtained from the peak values at $1/B=0.64$ and $0.077$ (1/T) is plotted as a function of temperature in Fig. 4(b). The cyclotron effective mass $m^*/m_0=0.54\pm0.02$ was obtained by fitting these data with the temperature reduction factor $R_T(B,T)=(2\pi^2m^*k_BT/{\hbar}eB)/\sinh(2\pi^2m^*k_BT/{\hbar}eB)$. ($m_0$ is the free electron mass.) This value deviates from $0.375$\cite{Nak13} and $0.394$\cite{Wil94} of split-off holes in bulk diamond. This may be due to band mixing caused by the gate-induced electric field\cite{Ohk75,Lan76}.

Now let us perform a more detailed analysis of the oscillations to obtain further insights into the hole gas. As shown above, there is a parallel conduction channel presumably due to the heavy-hole subband, in addition to the one that shows the SdH oscillations. Therefore, we performed an analysis based on $\sigma_{xx}$ and $\sigma_{xy}$ rather than $\rho_{xx}$ and $\rho_{yx}$. $\sigma_{xx}$ and $\sigma_{xy}$ can be obtained from $\rho_{xx}$ and $\rho_{yx}$ using $\sigma_{xx}=\rho_{xx}/(\rho_{xx}^2+\rho_{yx}^2)$ and $\sigma_{xy}=\rho_{yx}/(\rho_{xx}^2+\rho_{yx}^2)$ [Fig. 5(a) and 5(b)]. $\sigma_{xx}$ and $\sigma_{xy}$ fit the two-carrier model,
\begin{eqnarray}
\sigma_{xx}&=&\frac{en_1\mu_1}{1+(\mu_1B)^2}+\frac{en_2\mu_2}{1+(\mu_2B)^2},\\
\sigma_{xy}&=&\frac{en_1\mu_1^2B}{1+(\mu_1B)^2}+\frac{en_2\mu_2^2B}{1+(\mu_2B)^2}.
\end{eqnarray}
Here, the carrier density obtained from the SdH frequency is taken as $n_1$ for the split-off holes. $n_2$ for the heavy holes is assumed to be $5n_1$ as the calculation based on the Schr\"odinger-Poisson equations provides $n_2/n_1\approx5$ for the carrier density range corresponding to the experiment. By setting the mobilities $\mu_1$ and $\mu_2$ as the fitting parameters, satisfactory fits to both $\sigma_{xx}$ and $\sigma_{xy}$ are obtained. $\mu_1$ ($\mu_2$) obtained from $\sigma_{xx}$ and $\mu_1$ ($\mu_2$) obtained from $\sigma_{xy}$ are in good agreement [Fig. 5(c)], which suggests that the above assumption is reasonable. The satisfactory fit with the two-carrier model is consistent with the above picture that the carriers in a subband have a higher mobility and show SdH oscillations.

The oscillating part of $\sigma_{xx}$ and $\sigma_{xy}$ can be extracted by subtracting the background curves obtained by the two-carrier-model fitting [Figs. 5(d) and 5(e)]. The SdH oscillations in $\sigma_{xx}$ and $\sigma_{xy}$ of two-dimensional systems are theoretically described by\cite{Sho84,Isi86,Col89}
\begin{eqnarray}
\Delta\sigma_{xx}&=&2a_{xx}\frac{en_0\mu_0}{1+(\mu_0B)^2}\frac{2(\mu_0B)^2}{1+(\mu_0B)^2}R_T(B,T)R_D(B)R_S\cos{\left[2\pi\frac{B_F}{B}-\pi\right]},\\
\Delta\sigma_{xy}&=&-2a_{xy}\frac{en_0\mu_0^2B}{1+(\mu_0B)^2}\frac{3(\mu_0B)^2+1}{(\mu_0B)^2(1+(\mu_0B)^2)}R_T(B,T)R_D(B)R_S\cos{\left[2\pi\frac{B_F}{B}-\pi\right]},\\
R_T(B,T)&=&\frac{2\pi^2m^*k_BT/{\hbar}eB}{\sinh(2\pi^2m^*k_BT/{\hbar}eB)},\\
R_D(B)&=&\exp{\left[-\pi/(\mu_qB)\right]},\\
R_S&=&\cos{\left[\frac{\pi}{2}g^*\frac{m^*}{m_0}\frac{1}{\cos{(\theta)}}\right]}.
\end{eqnarray}
$\theta$ is the magnetic field orientation relative to the direction normal to the surface, $B=B_{a}\cos{(\theta)}$, where $B_{a}$ is the applied magnetic field. $\mu_0$ (${\equiv} e\tau_0/m^*$) is the transport mobility and $\mu_q$ (${\equiv} e\tau_q/m^*$) is the quantum mobility, each corresponding to the transport lifetime $\tau_0$ and quantum lifetime $\tau_q$. Generally, $\tau_0\ge\tau_q$ because all scattering processes equally contribute to the quantum lifetime, whereas small-angle scattering has a smaller effect on the transport lifetime\cite{Dav98}. We introduced temperature- and magnetic-field-independent coefficients $a_{xx}$ (${\textgreater}0$) and $a_{xy}$ (${\textgreater}0$); $a_{xx}=a_{xy}=1$ in the original theory. Note that the above equations are for holes, whereas those in Ref. [9] are for electrons; therefore, the sign of $\Delta\sigma_{xy}$ is opposite. The fitting of $\Delta\sigma_{xx}$ and $\Delta\sigma_{xy}$ with Eqs. (3)-(7) by setting $\mu_0$, $\mu_q$, $a_{xx}R_S$ (or $a_{xy}R_S$), and $B_F$ as fitting parameters led to a large uncertainty in the obtained parameter values. Instead, by assuming that $\mu_0$ and $\mu_q$ are identical to $\mu_1$ obtained by the two-carrier-model fitting and by setting $a_{xy}R_S$ and $B_F$ as fitting parameters, a reasonably good fit of $\Delta\sigma_{xy}$ was obtained, as shown in Fig. 5(e). [$a_{xy}R_S$ and $B_F$ obtained by the fitting are shown in Figs. 5(f) and 1(e).] Here, $R_T$ was set to unity because $R_T(T\to0)=1$ and the temperature was low enough. Changing $\mu_q$ within about 20\% did not cause any significant change in the fitting, but making $\mu_q$ twice or half as large clearly degraded the fitting. (See Fig. S4 of the Supplemental Material.) The parameter set obtained by the fitting to $\Delta\sigma_{xy}$ and $a_{xx}/a_{xy}=2.5$ reproduces $\Delta\sigma_{xx}$ [Fig. 5(d)]. Thus, the overall magnetic field dependences of $\sigma_{xx}$ and $\sigma_{xy}$ including the SdH oscillations are almost completely explained by the theory of SdH oscillations [Eqs. (3)-(7)] together with the two-carrier model [Eqs. (1) and (2)]. The same conclusion is reached for the experimental data on another device (see Fig. S3 of the Supplemental Material). The deviation of $a_{xx}$ and $a_{xy}$ from unity in our devices may be associated with localized states between Landau levels\cite{Col89}. (See the Supplemental Material.) 

\begin{figure}
\includegraphics[width=15truecm]{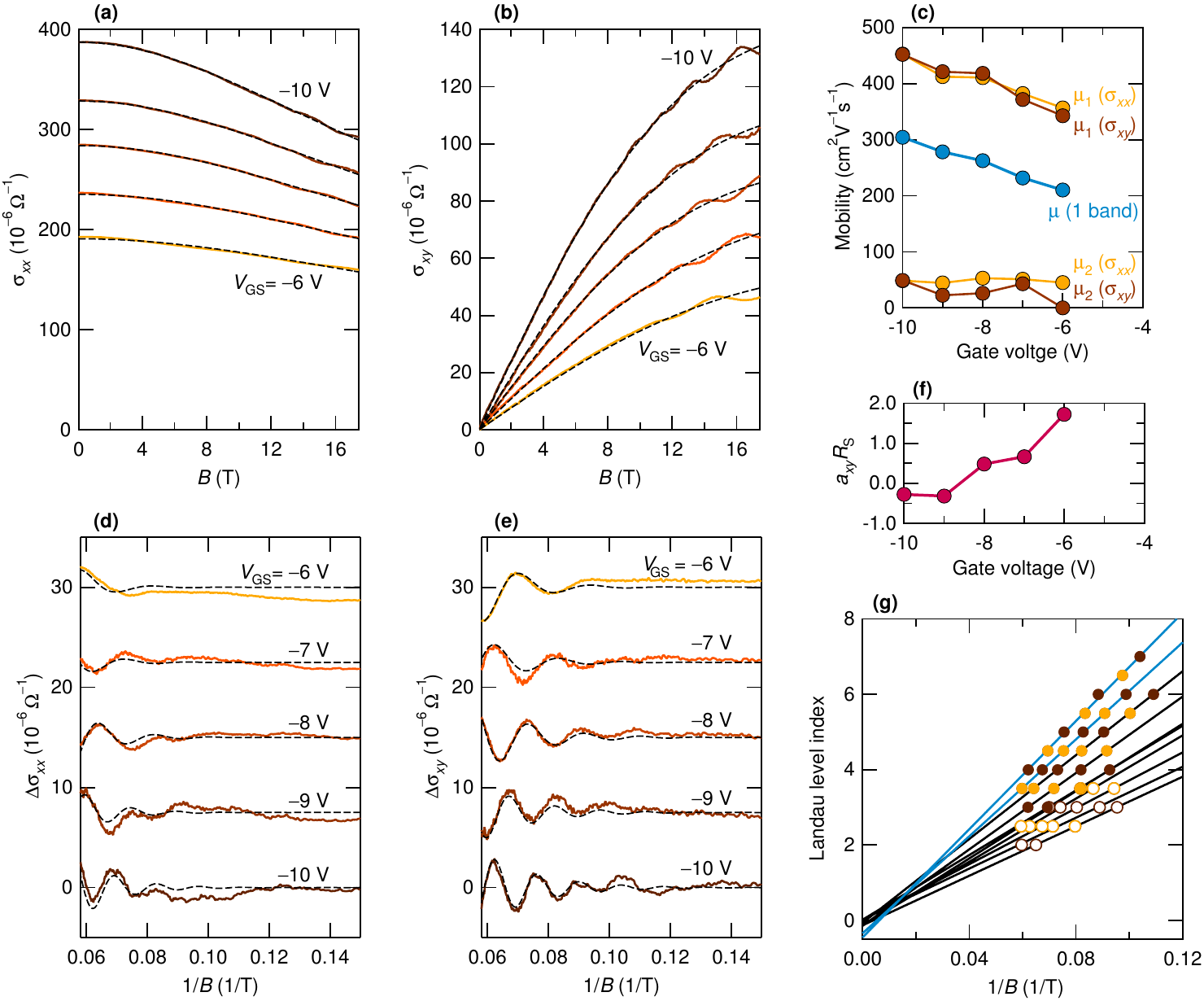}
\caption{\textbf{Detailed analysis of Shubnikov-de Haas oscillations.} (a) and (b) Magnetic field dependence of $\sigma_{xx}$ (a) and $\sigma_{xy}$ (b) of device S1 at 30 mK. The dashed lines are fits to the two-carrier model. (c) Mobilities of the two kinds of carrier obtained by the fitting. The figure also shows the mobility obtained by the single-carrier model using $\sigma_{xx}(B=0)$ and $\sigma_{xy}$ at low magnetic field. (d) and (e) The oscillating part of $\sigma_{xx}$ and $\sigma_{xy}$ obtained by subtracting the backgrounds [the dashed lines in (a) and (b)]. The curves are offset for clarity. The dashed lines are fits to the standard model of Shubnikov-de Haas oscillations. (see the text). (f) Gate voltage dependence of $a_{xy}R_S$ obtained by the fitting. (g) Fan diagram, showing the Landau indices plotted versus inverse magnetic fields at which $\Delta\sigma_{xy}$ has the maxima and minima; dark brown dots represent the maxima in $\Delta\sigma_{xy}$, corresponding to the minima in $\Delta\sigma_{xx}$, and light brown dots represent the minima in $\Delta\sigma_{xy}$, corresponding to the maxima in $\Delta\sigma_{xx}$. The closed dots are for device S1 at $V_\mathrm{GS}=-10$ to $-6$ V from top to bottom, and the open dots are for device S2 at $V_\mathrm{GS}=-4.5$ to $-3.5$ V from top to bottom. The lines are a linear fit, indicating the vertical-axis intercept close to -0.5 for the data of device S1 at $V_\mathrm{GS}=-10$ and $-9$ V (blue lines) and close to 0 for the other data (black lines).}
\end{figure}

The above results suggest that the quantum lifetime is close to the transport lifetime ($\mu_0/\mu_q {\approx} 1$) in the present system. The theoretical calculations indicate that $\mu_0/\mu_q$ can be larger than ten if the spatial separation $z_\mathrm{i}$ between the impurities and the carriers, or the Fermi wave vector $k_\mathrm{F}$, is large, whereas $\mu_0/\mu_q$ is close to unity for the opposite limit\cite{Sar85}. For example, $\mu_0/\mu_q{\ge}10$ if $z_\mathrm{i}$ is larger than ${\approx}1/k_F$ (for $z_\mathrm{i}q_\mathrm{TF}=0.5-8$), which is often the case in GaAs/Ga$_{1-x}$Al$_x$As heterostructures. ($q_\mathrm{TF}$ is the Thomas-Fermi screening constant.) The decrease in $z_\mathrm{i}$ or $k_\mathrm{F}$ leads to a decrease in $\mu_0/\mu_q$; $1{\le}\mu_0/\mu_q{\le}1.2$ for $z_\mathrm{i}$ smaller than ${\approx}0.1/k_F$ (for $z_\mathrm{i}q_\mathrm{TF}=0.5-8$). As $k_F=0.27-0.47$ nm$^{-1}$ for the carriers showing the SdH oscillations, the above results suggest that $z_\mathrm{i}$ is smaller than ${\approx}0.2-0.4$ nm in our devices. This is reasonable because the major scattering source in the present system is attributed to negatively charged impurities on the diamond surface\cite{Sas18}.

Interestingly, the sign of $R_S$ obtained above changes with the gate voltage. For device S1, $R_S$ was negative for $V_\mathrm{GS}=-9$ and $-10$ V, but positive for $V_\mathrm{GS}=-6$ to $-8$ V (Fig. 5f). $R_S$ was positive for all the gate voltages in the case of device S2, for which the SdH carrier density was lower than that for device S1 (Fig. S3j of the Supplemental Material). The change in the sign of $R_S$ means a change in the phase in the oscillation, which is also evidenced by the fan diagram [Fig. 5(g)]: the vertical-axis intercept is close to -0.5 for the linear fit to the data of device S1 at $V_\mathrm{GS}=-9$ and $-10$ V, but it is close to 0 for the other data. This result suggests that $g^*m^*/m_0{\textless}1$ (or $3{\textless}g^*m^*/m_0{\textless}5$) for $n_\mathrm{SdH}{\textless}2.6\times10^{12}$ cm$^{-2}$, and $1{\textless}g^*m^*/m_0{\textless}3$ (or $5{\textless}g^*m^*/m_0{\textless}7$) for $n_\mathrm{SdH}{\textgreater}3.2\times10^{12}$ cm$^{-2}$. The dependence of $m^*$ and/or $g^*$ on the gate voltage appears to be the origin of $R_S$ changing sign. Systematic measurements of $m^*$ and $g^*$ for different gate voltages will be required to clarify this point. $g^*$ can be determined from the detailed magnetic-field-orientation dependence of the oscillation; $g^*= n(m_0/m^*)\cos(\theta_C)$ ($n$=1, 3, 5 . . .), where $\theta_C$ is the angle at which the oscillation disappears due to the cancellation between the spin-up and spin-down oscillations. For such measurements, it will be important to improve the quality of devices by reducing the charged impurities on the diamond surface. Another approach to studying the carrier-density dependence of $g^*$ was recently taken on an ionic-liquid-gated diamond FET\cite{Akh19}, although it was in a carrier-density range an order of magnitude larger ($\ge 2\times10^{13}$ cm$^{-2}$) than in our FETs.

In summary, SdH oscillations were observed in the longitudinal and Hall resistivities of diamond FETs fabricated with monocrystalline $h$-BN as a gate dielectric. The frequency of the oscillation clearly depended on the gate voltage, and the corresponding carrier density could be understood in terms of the subband structure at the diamond surface. The SdH oscillations as well as the magnetic field dependence of $\sigma_{xx}$ and $\sigma_{xy}$ were quantitatively explained using a two-carrier model and the standard theory of the SdH oscillation for two-dimensional systems. This result indicates that a high-quality hole gas accumulated at the diamond surface. The ratio between the quantum and transport lifetimes is close to unity, which suggests that the charged impurities at the interface between diamond and $h$-BN are the major cause of carrier scattering. The high-quality hole gas demonstrated in this study has great potential for low-loss and high-speed electronics. Furthermore, it will open the door to a broad range of new research. For example, the low dielectric constant (=5.7$\epsilon_0$)\cite{Bha48} and heavy carrier effective masses\cite{Wil94,Nak13} of diamond may lead to strong carrier correlations and exotic quantum phenomena\cite{Kra03,Koz14}. The small spin-orbit interaction\cite{Win03} and low concentration of $^{13}$C with a nuclear spin\cite{Ter15} will be advantageous for exploring spintronic applications of high-mobility carriers in diamond. Furthermore, the possibility of electric-field-induced superconductivity in the low disorder limit is also interesting\cite{Phi98,Nakm13,Yam13,San17}. Thus, our results pave the way for studies of quantum transport in diamond as well as practical device applications. 

We thank H. Osato, E. Watanabe, and D. Tsuya for their technical support. We thank J. Inoue for useful discussions. We also thank T. Ando, S. Koizumi, T. Teraji, Y. Wakayama, and T. Nakayama for their kind support. This study was supported by Grants-in-Aid for Scientific Research (Grants No. 25287093, No. 26630139, No. 19H02605) and the ``Nanotechnology Platform Project'' of MEXT, Japan.

\bibliography{SasamaDiamondRev}

\newpage

\vspace{25truecm}
\textbf{Supplemental Material for "Quantum oscillations in diamond field effect transistors with a $h$-BN gate dielectric"}

\vspace{0.5truecm}
\subsection{Measurement setup.}

We performed measurements in a dilution refrigerator with an 18 T superconducting magnet. The gate bias was made using a source-measure unit (Keithley Instruments, 2400) through a filter or a function generator (Agilent Technologies, 33220A). The longitudinal and Hall resistivity were measured simultaneously using two lock-in amplifiers (Stanford Research Systems, SR830) at a frequency of 11.143 or 17.791 Hz. The ac voltage from the Sine Out output of one of the lock-in amplifiers was converted into current using a 10 M$\Omega$ series resistor and fed to the device. The current was in the range 10-50 nA. We checked the resistance values with a dc method using a function generator (Agilent Technologies, 33220A) for drain bias and voltage and current preamplifiers (Stanford Research Systems, SR560 and SR570). The longitudinal and Hall resistivity shown in Figs. 1c, 1d, S2, S3a, S3b, S4a, and S4b were obtained as $\rho_{xx}(B)=(\rho_{xx}(B)+\rho_{xx}(-B))/2$ and $\rho_{yx}(B)=(\rho_{yx}(B)-\rho_{yx}(-B))/2$.

\subsection{Calculation of subband structures at the diamond surface.}

The subband structure at the (111) surface of diamond for a given total sheet carrier density $n_\mathrm{2D}$ was calculated by solving the Schr\"odinger and Poisson equations (Eqs. 8-11) in a self-consistent manner\cite{Ham17}.
\begin{eqnarray}
\left[-\frac{\hbar^2}{2m_z^i}\frac{d^2}{dz^2}+e\phi(z)(+\Delta^\mathrm{SO})-E_n^i\right]\Psi_n^i(z)=0,\\
\frac{d^2\phi}{dz^2}=-\frac{1}{\epsilon}\left[eN_\mathrm{depl}+e\sum_{i,n}n_n^i|\Psi_n^i(z)|^2\right],\\
n_n^i=\frac{m_{//}^ik_BT}{\pi\hbar^2}\ln\left[1+\exp\left(\frac{E_F-E_n^i}{k_BT}\right)\right],\\
\sum_{i,n}n_n^i=n_\mathrm{2D}.
\end{eqnarray}
Similar calculations have been performed by Nebel et al.\cite{Neb04} and Edmonds et al.\cite{Edm10} Here, the energy is for holes; the sign of the energy is inverted from that of the usual description of band structure. The index $i$ represents the heavy hole (HH), light hole (LH), and split-off hole (SO). The calculation did not include the band mixing of the heavy, light, and split-off holes. We used the effective masses for the (111) surface derived from the Luttinger parameters\cite{Nak13}, $\gamma_1 = 2.670$, $\gamma_2 = -0.430$, and $\gamma_3 = 0.680$:
\begin{eqnarray}
m_z^\mathrm{HH}/m_0 = 1/(\gamma_1-2\gamma_3) = 0.763,\\
m_z^\mathrm{LH}/m_0 = 1/(\gamma_1+2\gamma_3) = 0.248,\\
m_z^\mathrm{SO}/m_0 = 1/\gamma_1 = 0.375,\\
m_{//}^\mathrm{HH}/m_0 = 1/(\gamma_1+\gamma_3) = 0.299,\\
m_{//}^\mathrm{LH}/m_0 = 1/(\gamma_1-\gamma_3) = 0.503,\\
m_{//}^\mathrm{SO}/m_0 = 1/\gamma_1 = 0.375.
\end{eqnarray}
The following values were used for the calculation: dielectric constant $\epsilon = 5.7\epsilon_0$,\cite{Bha48} spin-orbit splitting energy $\Delta^\mathrm{SO} = 6$ meV,\cite{Win03} the concentration of fixed space charges $N_\mathrm{depl} = N_\mathrm{D}-N_\mathrm{A} = 1.76\times10^{16}$ cm$^{-3}$ (0.1 ppm), and temperature $T = 0$ K. The Fermi level for large $z$ was set to $-E_\mathrm{D}$ below the conduction band bottom ($E_g-E_\mathrm{D}$ above the valence band top), where $E_\mathrm{D}=1.7$ eV is the donor (nitrogen) ionization energy and $E_{g} = 5.48$ eV is the bandgap of diamond.\cite{Squ06} The subband structures calculated for three different $n_\mathrm{2D}$ are shown in Fig. S5.

\subsection{Deviation of the SdH-oscillation amplitude from theory.}

The deviation of $a_{xx}$ and $a_{xy}$ from unity in our devices is very similar to the deviation from theory in GaAs/Ga$_{1-x}$Al$_x$As heterostructures\cite{Col89}, and probably has the same origin. The deviations of $a_{xx}$ and $a_{xy}$ from unity in our devices correspond to a deviation (increase) of the oscillation amplitude of $\Delta\rho_{yx}$ from theory especially for large $\mu_qB$, by a factor of $\approx2$ at most. The deviation of $\Delta\rho_{xx}$ is even smaller; it is less than $\approx30$\%. In the case of GaAs/Ga$_{1-x}$Al$_x$As heterostructures, SdH oscillations of $\rho_{xx}$ and $\rho_{xy}$ are well described by the theory with $a_{xx}{\approx}1$ and $a_{xy}{\approx}1$ when $\mu_qB$ is in the range 0.4-0.6.\cite{Col89} However, the amplitude of $\Delta\rho_{xy}$ becomes larger than the theoretical one for $\mu_qB{\textgreater}0.6$; the factor is ${\approx}2$ for $\mu_qB=1$. (For $\Delta\rho_{xx}$, agreement with theory is obtained even for $\mu_qB{\textgreater}0.6$.) The amplitude of $\Delta\rho_{yx}$ increased in our devices for $\mu_qB=0.5-1.1$, which is nearly the same range in which there is a similar increase in GaAs/Ga$_{1-x}$Al$_x$As heterostructures. It has been argued\cite{Col89} that the deviation from theory in GaAs/Ga$_{1-x}$Al$_x$As heterostructures is associated with localized states between Landau levels.

\newpage

\begin{figure}
\includegraphics[width=5.8truecm]{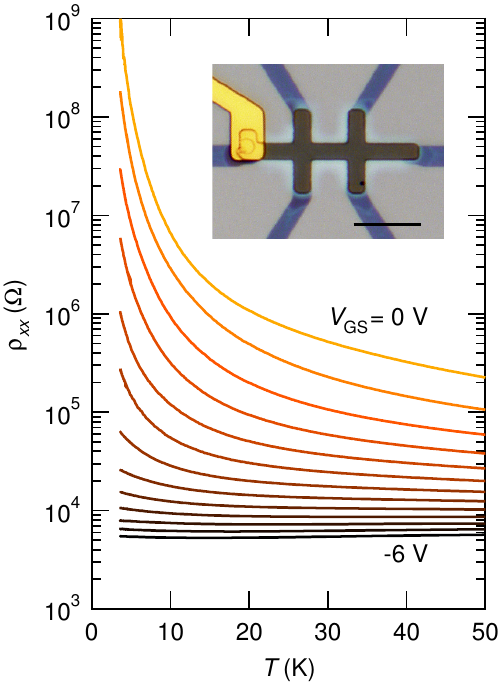}\\
\begin{flushleft}
{\footnotesize \textbf{Figure S1: Temperature dependence of longitudinal resistivity at different gate voltages.} Temperature dependence of the longitudinal resistivity of device S1 at different gate voltages (changed in 0.5-V increments). The inset shows an optical micrograph of device S1. Scale bar, 10 $\mu$m.}
\end{flushleft}
\end{figure}

\begin{figure}
\includegraphics[width=6.5truecm]{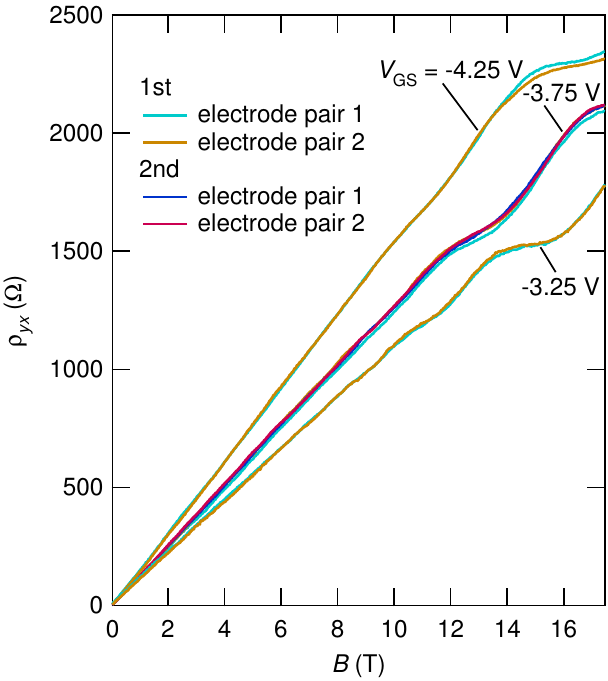}\\
\begin{flushleft}
{\footnotesize \textbf{Figure S2: Hall resistivity measured with different pairs of electrodes.} Hall resistivity of device S2 measured with different pairs of electrodes at $V_{GS}=-3.25$, $-3.75$, and $-4.25$ V at 30 mK. Measurements were performed twice for $V_{GS}=-3.75$ V; after the first measurement, the gate voltage was swept to -4.5 V, then to 0 V, and back to -3.75 V for the second measurement.}
\end{flushleft}
\end{figure}

\newpage

\begin{figure}
\includegraphics[width=15truecm]{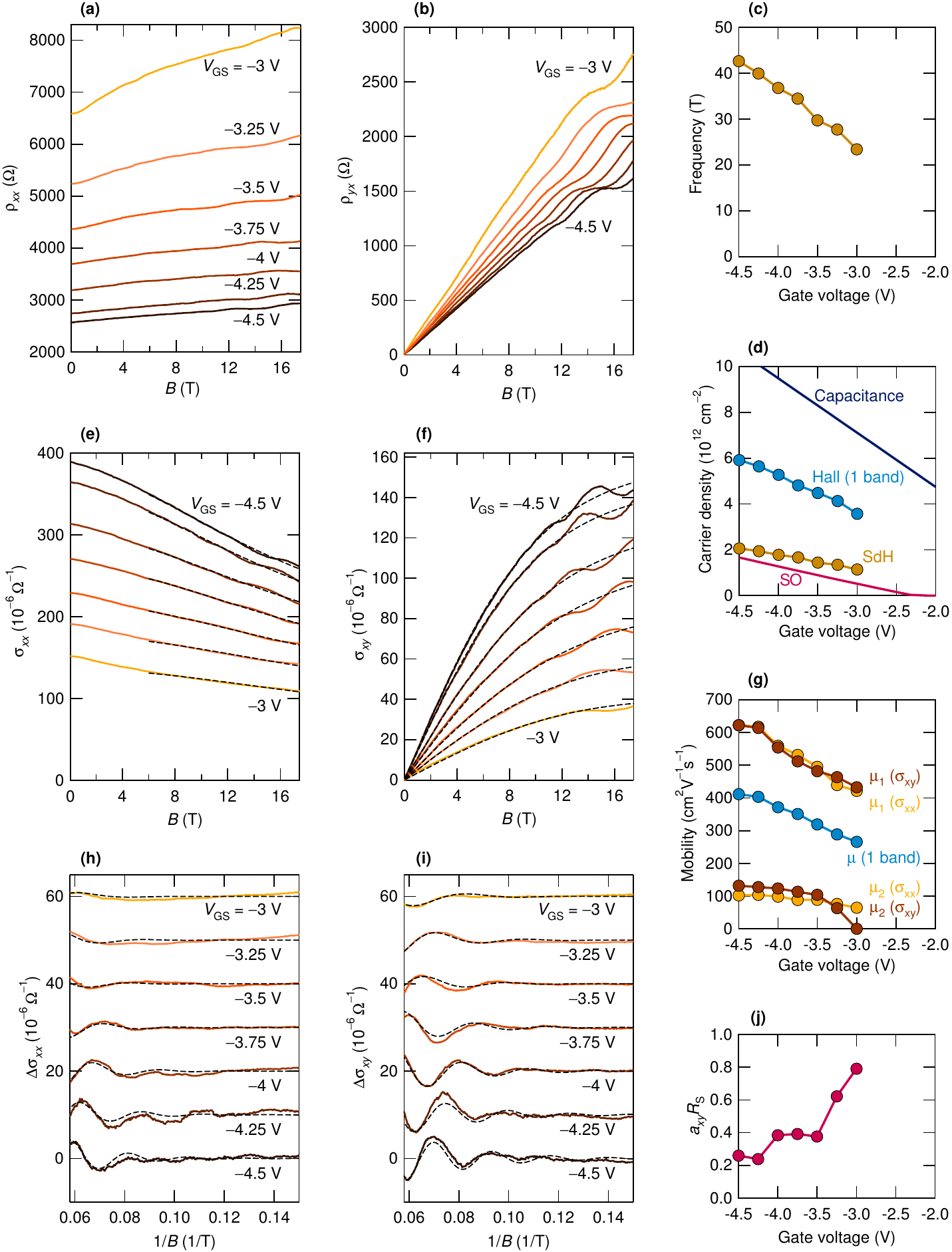}
\end{figure}

\newpage

\begin{figure}
\begin{flushleft}
{\footnotesize \textbf{Figure S3: Detailed analysis of SdH oscillations in device S2.} (a) and (b) Longitudinal (a; $\rho_{xx}$) and Hall (b; $\rho_{yx}$) resistivities at different gate voltages at 30 mK plotted as a function of magnetic field. (c) Frequency of SdH oscillation in $\sigma_{xy}[=\rho_{yx}/(\rho_{xx}^2+\rho_{yx}^2)$]. (d) Gate voltage dependence of the sheet carrier densities calculated from the SdH frequency and low-field Hall resistivity. The figure also shows the carrier density estimated from the gate capacitance and the simulated density of the holes occupying the split-off hole subband. (e) and (f) Magnetic field dependence of $\sigma_{xx}$ (e) and $\sigma_{xy}$ (f), calculated from $\rho_{xx}$ and $\rho_{yx}$ in (a) and (b). The dashed lines are fits to the two-carrier model. (g) Mobilities of the two kinds of carrier obtained by the fitting. The figure also shows the mobility obtained by the single-carrier model using $\sigma_{xx}(B=0)$ and $\sigma_{xy}$ at low magnetic field. (h) and (i) The oscillating part of $\sigma_{xx}$ and $\sigma_{xy}$ obtained by subtracting the backgrounds [the dashed lines in (e) and (f)]. The curves are offset for clarity. The dashed lines are fits to the standard model of Shubnikov-de Haas oscillations. $a_{xx}/a_{xy}=1.5$. (See the main text) (j) gate voltage dependence of $a_{xy}R_S$ obtained by the fitting.}
\end{flushleft}
\end{figure}

\newpage

\begin{figure}
\includegraphics[width=10truecm]{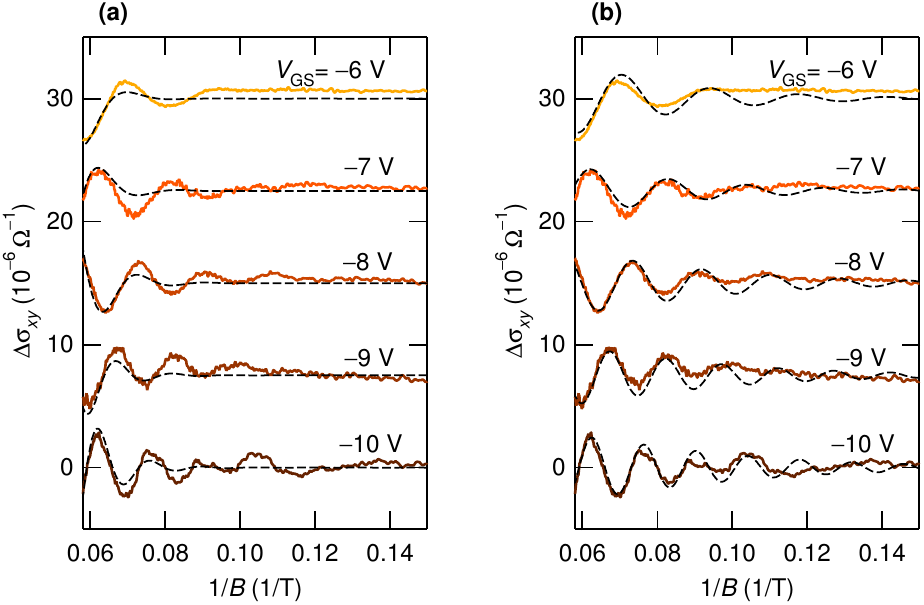}\\
\begin{flushleft}
{\footnotesize \textbf{Figure S4: Fitting to $\Delta\sigma_{xy}$ of device S1 by using different values of $\mu_q$.} (a) Result of fitting assuming that $\mu_q=0.5\mu_1$ and $\mu_0=\mu_1$. (b) Result of fitting assuming that $\mu_q=2\mu_1$ and $\mu_0=\mu_1$. A better fit is obtained when $\mu_q=\mu_0=\mu_1$. (Fig. 5e of the main text)}
\end{flushleft}
\end{figure}

\newpage

\begin{figure}
\includegraphics[width=15truecm]{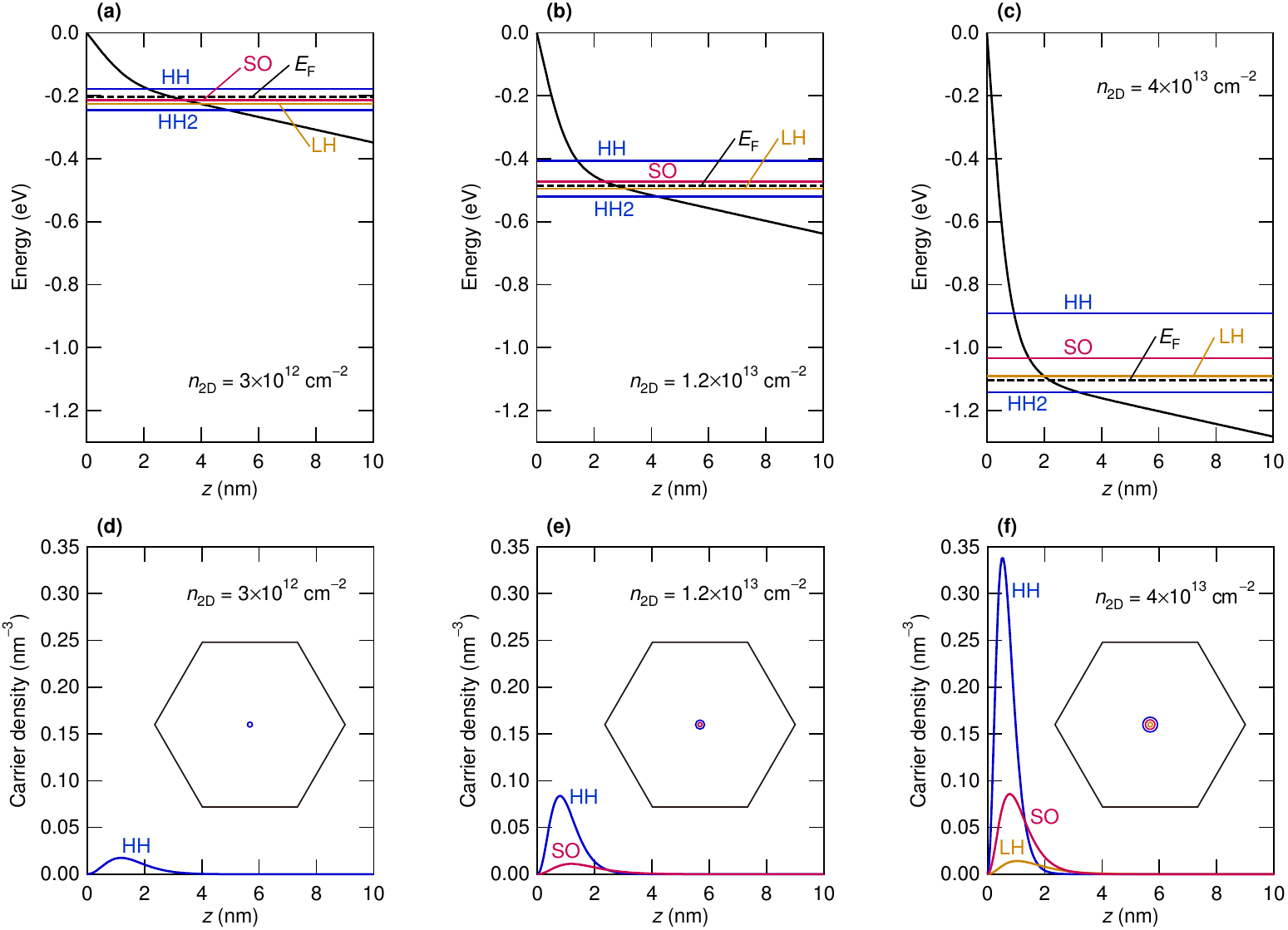}\\
\begin{flushleft}
{\footnotesize \textbf{Figure S5: Simulated subband structure at the diamond (111) surface for different total carrier densities.} Energies at the top of subbands [(a)-(c)] and carrier density profiles [(d)-(f)] obtained by solving the Schr\"odinger and Poisson equations in a self-consistent manner. $z$ is the depth from the diamond surface. HH, LH, and SO represent the lowest subbands of heavy, light and split-off holes, and HH2 represents the second subband of heavy holes. The total hole density is $4\times10^{12}$ cm$^{-2}$ [(a) and (d)], $1.2\times10^{13}$ cm$^{-2}$ [(b) and (e)], and $4\times10^{13}$ cm$^{-2}$ [(c) and (f)]. The inset in (d)-(f) shows the Fermi lines and two-dimensional Brillouin zone. The radius of the circumscribed circle of the Brillouin zone is $(4\pi/3)(\sqrt{2}/a)=16.61$ nm$^{-1}$, where $a=0.3567$ nm is the lattice constant of diamond\cite{Ree96}.}
\end{flushleft}
\end{figure}

\newpage

\begin{table}

{\footnotesize 
\begin{tabular}{|c|c|c|c|c|c|c|c|c|c|}
\hline
$V_\mathrm{GS}$ & $\mu_\mathrm{H}$ & $n_\mathrm{H}$ & $B_\mathrm{F}$ & $n_\mathrm{SdH}$ & $\mu_1^{xy}$ & $\mu_1^{xx}$ & $\mu_2^{xy}$ & $\mu_2^{xx}$ & $a_{xy}R_\mathrm{S}$ \\
{\scriptsize (V)} & {\scriptsize (cm$^2$V$^{-1}$s$^{-1}$)} & {\scriptsize ($10^{12}$ cm$^{-2}$)} & {\scriptsize (T$^{-1}$)} & {\scriptsize ($10^{12}$ cm$^{-2}$)} & {\scriptsize (cm$^2$V$^{-1}$s$^{-1}$)} & {\scriptsize (cm$^2$V$^{-1}$s$^{-1}$)} & {\scriptsize (cm$^2$V$^{-1}$s$^{-1}$)} & {\scriptsize (cm$^2$V$^{-1}$s$^{-1}$)} &  \\
\hline\hline
-6.00 & 210 & 5.72 & 42.2 & 2.04 & 343 & 357 & 0.139 & 45.0 & 1.73\\
\hline
-7.00 & 232 & 6.37 & 47.7 & 2.31 & 372 & 382 & 43.2 & 50.9 & 0.668\\
\hline
-8.00 & 262 & 6.78 & 54.3 & 2.63 & 418 & 411 & 26.1 & 52.8 & 0.481\\
\hline
-9.00 & 278 & 7.39 & 66.8 & 3.23 & 421 & 412 & 22.5 & 44.5 & -0.314\\
\hline
-10.00 & 304 & 7.94 & 72.0 & 3.48 & 452 & 453 & 49.2 & 48.4 & -0.277\\
\hline
\end{tabular}}
\begin{flushleft}
{\footnotesize \textbf{Table S1: Parameters for device S1.} $V_\mathrm{GS}$, gate voltage; $\mu_\mathrm{H}$ and $n_\mathrm{H}$, Hall mobility and Hall carrier density obtained from $\sigma_{xx}(B=0)$ and $\sigma_{xy}$ at low magnetic field by assuming a single-carrier model; $B_\mathrm{F}$, SdH frequency; $n_\mathrm{SdH}$, SdH carrier density; $\mu_1^{xy}$, $\mu_1^{xx}$, $\mu_2^{xy}$, and $\mu_2^{xx}$, mobilities obtained from $\sigma_{xx}(B)$ and $\sigma_{xy}(B)$ by the two-carrier-model fitting; $a_{xy}R_\mathrm{S}$, a coefficient times spin reduction factor. (see the main text)}
\end{flushleft}

{\footnotesize 
\begin{tabular}{|c|c|c|c|c|c|c|c|c|c|}
\hline
$V_\mathrm{GS}$ & $\mu_\mathrm{H}$ & $n_\mathrm{H}$ & $B_\mathrm{F}$ & $n_\mathrm{SdH}$ & $\mu_1^{xy}$ & $\mu_1^{xx}$ & $\mu_2^{xy}$ & $\mu_2^{xx}$ & $a_{xy}R_\mathrm{S}$ \\
{\scriptsize (V)} & {\scriptsize (cm$^2$V$^{-1}$s$^{-1}$)} & {\scriptsize ($10^{12}$ cm$^{-2}$)} & {\scriptsize (T$^{-1}$)} & {\scriptsize ($10^{12}$ cm$^{-2}$)} & {\scriptsize (cm$^2$V$^{-1}$s$^{-1}$)} & {\scriptsize (cm$^2$V$^{-1}$s$^{-1}$)} & {\scriptsize (cm$^2$V$^{-1}$s$^{-1}$)} & {\scriptsize (cm$^2$V$^{-1}$s$^{-1}$)} &  \\
\hline\hline
-3.00 & 266 & 3.57 & 23.4 & 1.13 & 433 & 422 & 0.363 & 64.8 & 0.791\\
\hline
-3.25 & 289 & 4.13 & 27.7 & 1.34 & 463 & 439 & 63.4 & 76.2 & 0.623\\
\hline
-3.50 & 319 & 4.48 & 29.7 & 1.44 & 482 & 495 & 104 & 88.9 & 0.376\\
\hline
-3.75 & 351 & 4.81 & 34.5 & 1.67 & 512 & 532 & 113 & 88.6 & 0.393\\
\hline
-4.00 & 372 & 5.27 & 36.8 & 1.78 & 555 & 559 & 124 & 99.0 & 0.385\\
\hline
-4.25 & 403 & 5.64 & 40.0 & 1.93 & 614 & 618 & 128 & 104 & 0.239\\
\hline
-4.50 & 411 & 5.91 & 42.7 & 2.06 & 622 & 623 & 132 & 102 & 0.260\\
\hline
\end{tabular}}
\begin{flushleft}
{\footnotesize \textbf{Table S2: Parameters for device S2.} The same parameters as those in Table S1 are shown. The mobility $\mu_1^{xy}=482$ cm$^2$V$^{-1}$s$^{-1}$ at $V_\mathrm{GS}=-3.5$ V corresponds to the lifetime $0.15$ ps by using the effective mass $m^*/m_0=0.54$.}
\end{flushleft}
\end{table}

\end{document}